%
%From misha@devil.tau.ac.il Sun Jan 19 14:44:25 1997
%Date: Sun, 19 Jan 1997 14:28:24 +0200 (IST)
%From: Misha Koslow <misha@devil.tau.ac.il>
%To: andelman@post.tau.ac.il

\font\twelvett=cmtt12 
\font\twelvebf=cmbx12 
\font\twelverm=cmr12 
\font\twelvei=cmmi12 
\font\twelveit=cmti12
\font\twelvesy=cmsy10 scaled\magstep1 
\font\twelveex=cmex10 at 12pt

\font\tnrm=cmr10 
\font\tni=cmmi10 
\font\tnbf=cmbx10 
\font\tnsy=cmsy10 
\font\tnex=cmex10

\font\eightrm=cmr8 
\font\eightbf=cmbx8 
\font\eighti=cmmi8 
\font\eightsy=cmsy8

\def\klein{ \def\rm{\fam0\twelverm}

\textfont0 =\twelverm \scriptfont0 =\tnrm
\scriptscriptfont0 =\eightrm
\textfont1 =\twelvei \scriptfont1 =\tni
\scriptscriptfont1 =\eighti
\textfont2 =\twelvesy \scriptfont2 =\tnsy
\scriptscriptfont2 =\eightsy
\textfont3 =\twelveex   \scriptfont3 =\tnex
\scriptscriptfont3 =\tnex

\textfont\itfam=\twelveit  \def\it{\fam\itfam\twelveit}
\textfont\ttfam=\twelvett  \def\tt{\fam\ttfam\twelvett}
\textfont\bffam=\twelvebf  \scriptfont\bffam=\tnbf
\scriptscriptfont\bffam=\eightbf
\def\bf{\fam\bffam\twelvebf}}

\vsize=24.5 true cm
\hsize=16 true cm

\hoffset=0. true cm
\voffset=-0.5 true cm

\baselineskip=0.8 true cm plus 0.2 true cm
\lineskiplimit=0.1 true cm
\lineskip=0.1 true cm minus 0.0 true cm
\raggedbottom

\parskip=0 true cm

\parindent=0.75 true cm

\nopagenumbers
\footline={\hss \tt -\ \folio \ - \hss}

\def\nz{\hfill\break}

\def\ob #1{\leavevmode\raise 0.5 ex\hbox{#1}}

\def\un #1{\leavevmode\lower 0.5 ex\hbox{#1}}

\tolerance=10000
\hbadness=10000

\binoppenalty10000
\relpenalty=10000

\def\etwas #1 #2 {\par\noindent \hangindent= #1\hangafter=0
\llap{#2\quad}\ignorespaces}

 \klein \rm \def\nz{\hfill\break}

\tolerance=10000 \hbadness=10000
\baselineskip=0.8 true cm plus 0.2 true cm
\binoppenalty10000 \relpenalty=10000 \rm  % %

\centerline{\bf Shape of phospholipid/surfactant mixed 
 micelles: cylinders or discs?} 
\centerline{\bf Theoretical analysis.}
\nz
\centerline{\bf M.M.Kozlov$^{1}$, D.Lichtenberg$^{1}$ and D.Andelman$^{2}$}
\nz\nz
$^{1}$ Department of Physiology and Pharmacology, Sackler School of Medicine,
Tel Aviv University, Ramat-Aviv 69978, Israel\nz
$^{2}$ School of Physics and Astronomy, Raymond and Beverly Sackler Faculty of
Exact Sciences, Tel Aviv University, Ramat-Aviv 69978, Israel
\nz\nz\nz 
\centerline{\bf Abstract.}
We develop a theoretical model for the solubilization of phospholipid 
bilayers by
micelle-forming surfactants. Cylindrical
micelles, disc-like micelles and spherical micelles are considered as
alternative resultant structures. The main question addressed is what
kind of micelles can be expected under various thermodynamical conditions. 
Our analysis is based on a theoretical model that accounts for Helfrich
energy of  
curvature of amphiphile monolayers and for the entropy of 
mixing of lipids and surfactants in mixed aggregates.

We conclude that for usual values of the elastic 
parameters of amphiphile monolayers cylindrical micelles are the most 
probable aggregates resulting from micellization of phospholipid by
surfactants. This conclusion is consistent with 
available experimental data. Conditions of formation
of disc-like and spherical micelles are also determined.
\eject
\centerline{\bf Introduction.}\nz\nz
Amphiphiles tend to self assemble in aqueous solutions, mostly due to the
hydrophobic effect$^1$. 
In the resultant aggregates, the amphiphilic molecules
are packed as monolayers, where their hydrophobic moieties are shielded
from contact with the external aqueous medium by the polar head groups.
Depending on their molecular structure and interactions, the amphiphiles
form aggregates of different shapes$^{2-4}$. Most of the
biological amphiphiles (phospholipids) self assemble in nearly flat
bilayer membranes, forming closed vesicles (liposomes). By contrast,
most of the
commonly used surfactants form micelles whose radii of curvature are close
to length of hydrocarbon chains $^5$.

While each of the pure compounds form in dilute solution 
aggregates of a particular type, 
mixtures
of lipids and surfactants self assemble in either
mixed liposomes or mixed micelles, depending on the composition
$^{5-7}$. 
Transition from mixed bilayers to mixed micelles upon
addition of surfactant to phospholipid vesicles, is commonly denoted as
solubilization of the liposomes.  
The resultant micelles were previously described as having either disc-like
(oblate ellipsoidal$^8$) or cylindrical shapes,$^5$ in apparent agreement with
dynamic light scattering data.

More recently, however, cryo-transmition electron
microscopy,$^{9,10}$ size-exclusion high
performance liquid chromatography,$^{11}$
small angle neutron scattering,$^{12}$ and re-evaluation of dynamic light
scattering data$^{13}$ indicated that   
in most cases solubilization of liposomes
results in formation of thread-like rather than  
disc-like micelles.

The model of disc-like micelles was supported by the idea
that the surfactant molecules form the rims of the 
 micelles where the amphiphile monolayers are strongly curved, while 
 the lipid molecules remain in flat parts of the discs. 
However, this qualitative 
consideration did not account for the entropy of mixing of the two components
in the micelle, which tends to distribute uniformly the molecules of the two
components over the whole surface of each micelle. The result of competition
of these two tendencies is not obvious and requires a detailed theoretical
analysis.

Based on the more recent experimental results, 
theoretical approaches that were developed to
describe the energetics and size distributions of mixed amphiphilic
aggregates $^{14,15}$ and to interpret the phase
diagrams of lipid-surfactant mixtures,$^{6-7}$ assumed 
that cylindrical and spherical micelles are the only possible
aggregates resulting 
from solubilization. The question
remains open whether disc-like micelles are indeed less favorable 
energetically than the cylindrical and spherical ones and if there
are conditions where 
solubilization can still result in formation of disc-like micelles.

The present work analyses and compares the conditions of
surfactant-induced phase transition of lipid bilayers into discoidal, 
cylindrical and spherical micelles. Our model, unlike 
alternative theoretical approaches$^{16}$, do not consider the detailed 
distribution of
microscopic interactions in amphiphile monolayers. Instead,
we describe a monolayer of amphiphiles by the more macroscopic Helfrich 
elastic model and account for the lipid/surfactant entropy of mixing. 
We show that the
shape of the micelles formed upon solubilization of liposomes is 
determined by a unique parameter that depends on the temperature and on the 
elastic characteristics of the monolayer as expressed by its bending
rigidity, the spontaneous curvatures of the two compounds 
and the Gaussian curvature modulus. 

Based on our calculations, we conclude that at
all reasonable values of this parameter the predicted shapes of the mixed
micelles are those of long cylinders, in agreement with the recent  
experimental data$^{9,10}$. Disc-like
micelles can only be expected for   
compounds whose Gaussian curvature modulus is of unusually high negative
value. Another possibility to obtain disc-like micelles
is to suppress the effects of the entropy of mixing by decreasing the 
temperature.
\nz\nz\nz
\centerline{\bf The model.}
We consider a ternary system of water, lipid and surfactant. The concentrations
of lipid and surfactant in water, denoted by $N_L$ and $N_D$, respectively,
are assumed to be much higher than the critical micelle concentrations (cmc).

We consider the following states of the aqueous solution of the amphiphiles in
terms of the most common shapes of the aggregates:
flat bilayers (liposomes, whose radius is taken as very large in comparison 
to the
bilayers thickness), cylindrical micelles, disc-like micelles, spherical 
micelles and mixtures of coexisting aggregates of these types. Since the aim 
of our work is to analyze qualitatively the main pathways of solubilization 
of bilayers we do
not consider the more complicated architecture of intermediate 
aggregates$^{17}$ such as 
ellipsoidal micelles, hyperbolic and mesh structures$^{17-20}$.  

To characterize the composition of the system we use the area
fraction occupied by surfactant
\nz
$$\phi = {{a_D N_D}\over{a_D N_D + a_L N_L}}\eqno(1)$$
\nz  
where $a_D$ and $a_L$ are the molecular areas of the 
surfactant and lipid, respectively, at the monolayer plane. It is assumed
that those specific areas do not differ for aggregates of various types. 

In our model, the pure lipid system 
($\phi = 0$) preferentially forms liposomes, while addition of the surfactant
results in transition of liposomes into micelles of one of the shapes
mentioned above. To analyze the resulting structures, we determine 
for each of the possible aggregates mentioned
above the free energy as a function of composition. 
Comparing those free energies,
we find for each composition the state of the lowest free energy, i.e. 
the equilibrium structure. 

The obvious difference between the structures of aggregates 
is the curvature of their monolayers. Therefore, as a 
first contribution to the free energy (per unit area) of the monolayer, $u$,
we consider the Helfrich energy of bending  
$u_b$.$^{21}$  The other major contribution is that of the entropy of mixing of
the two components in the monolayer, $s$.$^6$  Hence,   
\nz
$$u = u_b - T\cdot s$$
\nz
We neglect the translational entropy of aggregates and the entropy of 
polydispersity of the micellar sizes$^{14}$ since 
the related contribution to the free energy can be shown to correct
only slightly the criteria of solubilization of the bilayers.

The energy of bending (per unit area) of the monolayer is 
\nz
$$ u_b = {1\over 2}\kappa (c_m + c_p -c_0)^2 + \bar\kappa c_m c_p\eqno(2)$$
\nz
where $c_m$ and $c_p$ are the principal curvatures of the bent surface
.$^{21}$ The material properties of the monolayer 
that determine the bending energy (2) are
the bending rigidity $\kappa$, the Gaussian curvature modulus $\bar\kappa$
and the spontaneous curvature $c_0$. 

The model (2) has been originally 
formulated for membrane shapes that deviates only slightly from flat surface
.$^{21}$ Extension of this model to the cases of 
strongly curved micelles can be justified 
by recent analysis of the elastic properties
of monolayers of inverted hexagonal phases, whose curvature is comparable 
(but opposite in sign) to the curvature of cylindrical micelles.$^{22,23}$  

For the sake of simplicity we will assume  that the moduli $\kappa$ and
$\bar\kappa$ do not depend on composition. While several  
theoretical models
predict different characters of such dependencies,$^{24,25}$ 
a more recent numerical calculation$^{26}$ 
revealed a weak dependence of $\kappa$ on the 
composition of a mixed membrane. Moreover, an experimental determination of
the bending rigidity of mixed monolayers of H$_{II}$ phases did not show any
pronounced changes of $\kappa$ with composition.$^{23}$

By contrast, the spontaneous curvature $c_0$ depends strongly on the monolayer 
composition. Although this dependence may be rather complicated,$^{15, 27}$ 
we will assume that $c_0$ is the area-weighted average
of the spontaneous curvatures of pure lipid $c_L$ and pure surfactant $c_D$,
\nz
$$c_0 =(1-\phi) c_L + \phi c_D \eqno(3)$$
\nz
This assumption is supported by numerical calculations$^{26}$ that
showed that the spontaneous curvature of a monolayer consisting of surfactants
with different chain lengths is a linear function of $\phi$ 
over a wide range of 
compositions. Furthermore, for mixed monolayers of 
H$_{II}$ phases it has been shown experimentally$^{23}$ 
that $c_0$ depends linearly on $\phi$. 

To account for the entropy of mixing (per unit area) of the two components 
we will use an approximate expression for ideal mixing $^6$ 
\nz
$$  {s\over k_B} 
= -{1\over a_L} [\phi\log\phi + (1-\phi)\log (1-\phi)] \eqno(4) $$
\nz
The following analysis depends mainly on the behavior of the entropy $s$ near
its minimal value at $\phi = {1\over 2}$. Within this range (4) 
can be substituted by a simpler equation
\nz
$$  {1\over k_B}\cdot s  =  -{1\over a_L} [2 \phi (\phi - 1) - 0.2] \eqno(5)$$
\nz
Eq.(5) in fact approximates (4) with a good accuracy for any value of 
$\phi$ within the range $ 0.2 < \phi < 0.8$, which covers the whole range
of "solubilizing ratios" in any lipid/surfactant mixture studied thus far.

We will assume below that the characteristic radius of the curved parts of
amphiphile monolayers is equal $\rho$
for all kinds of aggregates. It is convenient to express all the variables
in dimensionless form. The dimensionless free energy per unit area $f$,
the dimensionless curvature $\zeta$, and the dimensionless temperature
$\eta$ are defined as 

$ f = {2\rho^2\over\kappa}\cdot u$,
 $\zeta = \rho c$, and
$\eta = {4 k_B T\rho^2\over{\kappa a_L}}$. 

The dimensionless
energy per unit area of the monolayer is
\nz
$$f = (\zeta_m + \zeta_p - \zeta_0)^2 + 
2\cdot{\bar\kappa\over\kappa}\cdot\zeta_m\cdot\zeta_p + 
\eta\cdot\phi (\phi - 1) + const \eqno(6)$$
\nz
where $\zeta_0$ depends on $\phi$ according to (3).

Equations (1)-(6) are the basis for the determination of the 
 free energies of different micellar states and 
liposomes (flat bilayers) as functions of their compositions.
\nz\nz
\centerline{\bf Free energies.}\nz\nz
{\it Disc-like micelles.}\nz 
The shape of a disc-like micelle is assumed
to consist of a flat central part with radius $R$ and a rim formed by a
strongly curved monolayer, whose meridional principal curvature is
$1/\rho$ (Fig.1). The areas of the rim and the flat
part are denoted as $A_e$ and $A_f$, respectively. 
The geometrical characteristics of the monolayer forming the rim are
considered in details in Appendix A.

To calculate the free energy of a disc-like micelle $f_D$ as a function of its 
composition $\phi_D$, we take into account the partitioning of
the surfactant between the flat part and the rim (Appendix B) while determining 
the area-weighted average of the
energy (6). Denoting the dimensionless total curvature in the
rim as $J = (c_m + c_p)\rho$, we obtain for the free energy
(per unit area)
\nz
$$ f_D = [(\Delta\zeta)^2 + \eta]\phi^2_D +
(2\zeta_L\Delta\zeta - \eta - 2{A_e\over A_t}\Delta\zeta <J>)\phi_D$$
$$+ {A_e\over A_t}\Biggl[ {\eta\over{\eta+(\Delta\zeta)^2}}<J^2> +
{(\Delta\zeta)^2\over{\eta+(\Delta\zeta)^2}}<J>^2\Biggr] -$$
$$ - {A_e A_f\over A_t^2}{(\Delta\zeta)^2\over{\eta+(\Delta\zeta)^2}}<J>^2
- 2 {A_e\over A_t}\zeta_L<J> + \zeta_L^2 +
8\pi{\bar\kappa\over\kappa}{\rho^2\over A_t}
\eqno(7)$$
\nz
where $<J>$ and $<J^2>$ are the values of the total curvature and its square,
averaged over the curved area forming the rim,
and $\Delta\zeta = \zeta_D - \zeta_L$ is the difference between the
spontaneous curvatures of surfactant and lipid expressed in dimensionless form.
\nz\nz
{\it Cylindrical micelles.}\nz 
We assume that all the cylindrical micelles have the
same shape consisting of a cylindrical 
part of a length $l$ with two semi-spherical
caps at the ends (Fig.2). The radius of the cross section 
of the cylinder and of the hemispheres 
is assumed to be equal to
$\rho$. The areas of the cylindrical part and the caps 
will be denoted as $A_l$ and $A_h$, respectively, and the total area 
is $A_t = A_l + A_h$.   

To derive the energy of the phase of cylindrical micelles (Appendix C) we
first determine the partitioning of the surfactant
between the less curved cylindrical body and more curved semi-spherical 
caps.  Subsequently, we
average the energy (6) over the area of one cylindrical micelle, including 
its body and two caps. As a result we obtain the free energy per unit area
as a function of the micellar composition $\phi_c$
\nz
$$f_c = [(\Delta\zeta)^2 + \eta]\phi^2_c +
(2\zeta_L\Delta\zeta - \eta - 2\Delta\zeta)\phi_c 
- 2 \Delta \zeta\phi_c {A_h\over A_t}$$
$$ + (\zeta_L-1)^2 - 
{(\Delta\zeta)^2\over{[(\Delta\zeta)^2 + \eta]}}{{A_h A_l}\over A_t^2}+
(3-2\zeta_L){A_h\over A_t} + 8\pi{\bar\kappa\over\kappa}{\rho^2\over A_t}
\eqno(8)$$
\nz\nz
{\it Spherical micelles.}\nz
The energy per unit area of spherical micelles of radius $\rho$ derived
from (6) as
a function of their composition $\phi_s$ is
\nz
$$ f_s =[(\Delta\zeta)^2 + \eta]\phi^2_s
+ (2\zeta_L\Delta\zeta - \eta - 4\Delta\zeta)\phi_s + 4
- 4\zeta_L + 2 {\bar\kappa\over\kappa} + \zeta_L^2 \eqno(9)$$
\nz\nz
{\it Flat bilayer of a liposome.}\nz
The energy per unit area of a monolayer of a flat bilayer
$f_b$ as a function of its composition $\phi_b$ is given by 
\nz
$$f_b =[(\Delta\zeta)^2 + \eta]\phi^2_b
+ (2\zeta_L\Delta\zeta - \eta)\phi_b + \zeta_L^2\eqno(10)$$
\nz
\nz
{\it Liposomes to Micelles Transition.}\nz
We analyze the transition from liposomes to micelles by Gibbs graphical 
method, illustrated in Fig.3 for one particular set of 
parameters of the system. The energies of the
pure phases (7)-(10) are convex functions of composition as presented 
by the  curves $b$, $c$ and $d$  for the 
bilayers (liposomes), cylindrical micelles and disc-like micelles,
respectively. The energies of mixtures of coexisting
phases are presented by common tangents
(dashed lines on Fig.3).  Compositions $\phi^*$
at the points where the common tangents touch the energy of bilayers
(curve b) indicate phase transitions.
We denote the compositions 
determining the transitions of liposomes to cylindrical, 
disc-like and spherical micelles, 
by
$\phi^*(b\rightarrow c)$, $\phi^*(b\rightarrow d)$ and
$\phi^*(b\rightarrow s)$
respectively.

The micelles for which the composition of transition $\phi^*$ has the
lowest value are expected to be formed upon solubilization.

The details on determination of the compositions of transition of the basis 
of Eqs.(7)-(10) are described in Appendix D. 

We obtain for the transitions from bilayers to cylindrical micelles:
\nz
$$ {{\eta + (\Delta\zeta)^2}\over\eta}\cdot 
\Biggl[\zeta_L + \Delta\zeta\cdot\phi^*(b\rightarrow c)\Biggr] =
{1\over 2} + {A_h\over {A_h + A_t}} + 
{\bar\kappa\over\kappa}\cdot{{\eta + (\Delta\zeta)^2}\over\eta}\cdot
{{4\pi\rho^2}\over{A_h+A_t}}\eqno(11)$$
\nz
for the transitions from bilayers to disc-like micelles:
\nz
$${{\eta + (\Delta\zeta)^2}\over\eta}\cdot
\Biggl[\zeta_L + \Delta\zeta\cdot\phi^*(b\rightarrow d)\Biggr] =
{1\over 2}{<J^2>\over<J>} +
{\bar\kappa\over\kappa}\cdot{{\eta + (\Delta\zeta)^2}\over\eta}\cdot
{{4\pi\rho^2}\over{A_e<J>}}\eqno(12)$$
\nz
and for the transition from bilayers to spherical micelles:
\nz
$${{\eta + (\Delta\zeta)^2}\over\eta}\cdot
\Biggl[\zeta_L + \Delta\zeta\cdot\phi^*(b\rightarrow s)\Biggr] =
1 + {1\over 2}{\bar\kappa\over\kappa}\cdot{{\eta + (\Delta\zeta)^2}\over\eta}
\eqno(13)$$
\nz
\nz
\centerline{\bf Criteria for the shape of micelles.}\nz\nz 
To compare  the critical compositions given by Eqs.(11)-(13), we use the
expressions for $A_e$, $<J^2>$ and $<J>$ derived in Appendix A. 

It follows from (11)-(13) that the
type of the formed micellar phase is controlled by a unique parameter
\nz
$$ \lambda = 
{\bar\kappa\over\kappa}\cdot {{\eta + (\Delta\zeta)^2}\over\eta}\eqno(14)$$
\nz
To illustrate it, we show in Fig.4 (A to C) the dependence of the critical 
composition for the transitions of the bilayer into cylindrical micelles
$\phi^*(b\rightarrow c )$ and 
into disc-like micelles $\phi^*(b\rightarrow d)$ as functions of the
micellar area $A_t$ for different values of $\lambda$ 
($\lambda = - 0.3$ in panel A, 
$\lambda = - 0.6$ in panel B,
$\lambda = -2.$ in panel C). 

For comparison, we show 
on the same  figure the
critical composition for the transition into spherical micelles  
$\phi^*(b\rightarrow s)$ 
presented as a constant dashed line as the size of micelles of this type is
fixed. 
As obvious from Figs.4, 
for $\lambda = -0.3$ (Fig.4A) the lowest critical
composition corresponds to the transition into long cylindrical micelles;
for $\lambda = -0.6$ (Fig.4B) 
the lowest critical
composition corresponding to the transitions into the disc-like micelles
of finite area;
and, finally, in the case $\lambda = -2.$ (Fig.4C) the lowest critical
composition is equal zero and corresponds to the transition into spherical 
micelles.

Detailed analysis of the model shows that there are two values of the
parameter $\lambda$, equal to $\lambda_1 = - 1/2$ and  $\lambda_2 = -2$ that 
separate the solubilization of bilayer into three different regimes.
For
\nz
$$\lambda > -1/2 \eqno(15)$$
\nz
the phase transition results in formation of long cylindrical
micelles;\nz 
for
\nz
$$ - 2 <\lambda < -1/2\eqno(16) $$
\nz
the transition leads to formation of
disc-like micelles of a finite radius,\nz 
while for 
\nz
$$\lambda < -2\eqno(17)$$
\nz
the bilayer transforms into spherical micelles.
This is illustrated in Fig.5, which shows the type 
of phase transition for different values of the parameter $\lambda$.
\nz\nz
\centerline{\bf Discussion.}\nz\nz
We have shown that the type of micelles resulting from solubilization
of bilayers is determined by the value of the parameter $\lambda$ (14).
This parameter depends on 
the difference of spontaneous curvatures of surfactant and lipid
$\Delta c = c_D - c_L$, the bending rigidity $\kappa$, the
Gaussian curvature modulus $\bar\kappa$, and the temperature $T$
\nz
$$ \lambda = 
{\bar\kappa\over\kappa}\cdot {{k_B T + 
{1\over 4}\kappa\cdot a_L\cdot (\Delta c)^2 }
\over{k_B T}}\eqno(18)$$
\nz
To understand qualitatively these results, let us recall that according to
the model of membrane elasticity $^{21}$ the
Gaussian curvature modulus
$\bar\kappa$, which 
controls the tendency of
the monolayer to change its topology,
can be either positive or negative. Negative
values of $\bar\kappa$ favor the division of each closed monolayer into as
large as  
possible number of separated closed monolayers, whereas positive values
of $\bar\kappa$ result in the opposite tendency, i.e. in recombination of 
separated
membranes into a single one. Hence, negative values of $\bar\kappa$
(yielding negative values of $\lambda$) prefer
large number of relatively small disc-like or spherical micelles 
with respect to a fewer long cylindrical micelles.
An opposite tendency competing with the effects of the negative $\bar\kappa$ 
relates to the energy of bending. This energy is controlled by the bending 
modulus $\kappa$, the difference in spontaneous curvatures of the components 
$\Delta c$ 
and the effectiveness of repartitioning of the surfactant between the parts
of the monolayers with different curvature, which, in turn is determined by
the temperature $T$.
The competition between these tendencies is
expressed by  
the parameter $\lambda$ (18) and criteria (15)-(17).  

Most of the parameters needed to estimate $\lambda$ are known or can 
be estimated with
reasonable accuracy. We will assume an area per lipid molecule
$a_L = 0.6 nm^2$, a radius of curvature of micelles $\rho = 1.5 nm$, a
bending modulus of the monolayer $\kappa = 10 k_B T$ (at room temperatures), 
a spontaneous
curvature of lipid $c_L = 0$, and a spontaneous curvature of the
surfactant $c_D = 1 /\rho$.

In the lack of a reliable experimental value for the modulus of 
Gaussian curvature of the monolayer $\bar\kappa$, we relate to the
theoretical prediction of $\bar\kappa$, as derived from recently developed 
models$^{24,28}$ which  
predict  $\bar\kappa$ to be negative 
and quite small in its absolute value.

With these estimates we can re-express
criteria (15)-(17) in terms of
the value of the Gaussian curvature modulus $\bar\kappa$. In
particular,  
formation of disc-like micelles occurs only when
\nz
$$ {\bar\kappa\over\kappa} < - 0.2\eqno(19) $$
\nz
Numerical calculations$^{24}$ performed on lipids and surfactants with usual 
characteristics give larger values of $\bar\kappa$
than required by (19). Therefore, formation of disc-like micelles
seems to be a very rare event occurring only for lipid/surfactant
mixtures with  
unusual properties. This prediction is in agreement 
with recent experimental results. However, we note that  
molecules with highly negative
$\bar\kappa$, which satisfy (19), can exist. For such compounds, formation
of disc-like micelles  should be expected.$^{29}$

In conclusion, present model 
provides the basis for understanding the surfactant-induced transformation
of bilayers into various types of mixed micelles. Nonetheless, predictions
based on this model
should be considered as
qualitative rather than quantitative ones. Further development of the model
requires more detailed experimental information on the elastic properties
of mixed amphiphilic monolayers.

\nz\nz
{\bf Acknowledgment.}
We would like to thank R.Granek, W.Helfrich and S.Safran for stimulating 
discussions. Support from the German-Israeli Foundation (GIF)
under grant No.I-0197 and the Deutsche Forschungsgemeinschaft through SFB 312
is gratefully acknowledged.
\eject
\centerline{\bf References.}
\item{$^1$} Tanford C., {\it The Hydrophobic Effect - Formation of Micelles
and Biological Membranes} 2nd.ed., John Wiley and Sons, 1980.

\item{$^2$} Israelachvili, J.N.; Mitchell,D.J.; Ninham,B.W.
{\it J.Chem.Soc.Faraday Trans.II} {\bf 1976}, {\it 72}, 1521.

\item{$^3$} Israelachvili,J.N.;
Marcelja,S.; Horn,R.G. 
{\it Q.Rev.Biophys.}{\bf 1980}, {\it 13}, 121. 

\item{$^4$} Gelbart W.M.; Roux,D.; and Ben-Shaul,A. (Editors). {\it Modern
Ideas and Problems in Amphiphilic Sciences.} Springer, 1993.

\item{$^5$} Lichtenberg D. Liposomes as a model for solubilization
and reconstitution of membranes. In: {\it Handbook of Nonmedical
Applications of
Liposomes}. v.II, Y.Barenholz and D.D.Lasic eds., CRC Press, Boca Raton New
York London Tokyo, 1996.

\item{$^6$} Andelman D.; Kozlov,M.M.; Helfrich,W. 
{\it Europhys.Lett.}{\bf 1994}, {\it 25}, 231.

\item{$^7$} Fattal,D.R.; Andelman,D.; Ben-Shaul,A.  
{\it Langmuir.}{\bf 1995}, {\it 11}, 1154. 

\item{$^8$} Schurtenberger,P.; Mazer,N.A.; Kanzig,W.  
{\it J.Phys.Chem.}{\bf 1985}{\it 89}, 1042.   

\item{$^9$} Walter,A.; Vinson,P.K.; Kaplun,A.; Talmon,Y. 
{\it Biophys.J.}{\bf 1991}, {\it 60}, 1315.

\item{$^{10}$} Vinson,P.K.; Talmon,Y.; Walter,A. 
{\it Biophys.J.}{\bf 1989}, {\it 56}, 669.

Edwards,K; Almgren,M.; Bellare,J.; Brown,W. {\it Langmuir} {\bf 1989}, 
{\it 5}, 473.

Edwards,K; Almgren,M. {\it J.Colloid Interface.Sci.} {\bf 1991}, {\it 147}, 1  

\item{$^{11}$} Nichols,J.W.; Ozarowski,J.{\it Biochemistry} {\bf 1990}, 
{\it 29}, 4600.

\item{$^{12}$} Pedersen,J.S.; Egelhaaf,S.U.; Schurtenberger,P. 
{\it J.Phys.Chem.} 
{\bf 1995}, {\it 99}, 1299  

Hjelm,R.P.; Alkan,M.H.; Thiyagarajan,P.  
 {\it Mol.Cryst.Liq.Cryst.} {\bf 1990}, {\it 180A}, 155.

\item{$^{13}$} Egelhaaf,S.U.; Schurtenberger,P. {\it J.Phys.Chem.} 
{\bf 1994}, {\it 98}, 8560

\item{$^{14}$} Ben-Shaul,A.; Rorman,D.H.; Hartland,G.V.; Gelbart,W.M.  
{\it J.Phys.Chem.} {\bf1986}, {\it 90}, 5277.

\item{$^{15}$} Szleifer,I.; Ben-Shaul,A.; Gelbart,W.M. 
{\it J.Chem.Phys.} {\bf 1987}, {\it 86}, 7094.

\item{$^{16}$} Zoeller,N.J; Blankschtein,D.  
{\it Ind.Eng.Chem.Res.} {\bf 1995}, {\it 34}, 4150.

\item{$^{17}$} Kekicheff,P.; Tiddy,G.J.T.  
{\it J.Phys.Chem.} {\bf 1989}, {\it 93}, 2520.

\item{$^{18}$} Hyde,S.T. 
{\it Pure and Appl. Chem.} {\bf 1992}, {\it 64}, 1617.

\item{$^{19}$} Hyde,S.T.  
{\it Colloque de Physique} {\bf 1990}, {\it C7}, 209.

\item{$^{20}$} Fredrickson,G.H. {\it Macromolecules} {\bf 1991},
{\it 24}, 3456.

\item{$^{21}$} Helfrich,W. {\it Z.Naturforsch.} {\bf 1973},
{\it 28(c)}, 693. 

\item{$^{22}$} Kozlov,M.M.; Leikin,S.; Rand,R.P.{\it Bioph.J.}
{\bf 1994}, {\it 67}, 1603.

\item{$^{23}$} Leikin,S.; Kozlov,M.M.; Fuller,N.L.; Rand,R.P. 1996. 
{\it Biophys.J.} {\bf 1996}, {\it 71}, 2623. 

\item{$^{24}$} Szleifer,I.; Kramer,D.; Ben-Shaul,A.; Gelbart,W.M.; 
Safran,S.A. {\it J.Chem.Phys.} {\bf 1990}, {\it 92}, 6800.

\item{$^{25}$} Dan,N.; Safran,S.A. {\it Europhys.Lett.} {\bf 1993}, 
{\it 21}, 975.\nz
Dan,N.; Safran,S.A.{\it Macromolecules} {\bf 1994}, {\it 27}, 5766.

\item{$^{26}$} May,S.; Ben-Shaul,A. 
{\it J.Phys.Chem.} {\bf 1995}, {\it 103}, 3839.

\item{$^{27}$} Kozlov,M.M.; Helfrich,W. {\it Langmuir} 
{\bf 1992}, {\it 8}, 2792.  

\item{$^{28}$} Safran,S.A. {\it Statistical Thermodynamics of Surfaces,
Interfaces, and Membranes.} Addison-Wesley Publ.Comp., 1994.

\item{$^{29}$} Granek,R.; Gelbart,W.M.; Bohbot,Y.; Ben-Shaul,A. 
{\it J.Chem.Phys.} {\bf 1994}, {\it 101}, 4331.
\eject
\nz\nz
\centerline{\bf Appendix A: Geometrical model of disc-like micelle.}\nz
We model the rim of a disc-like micelle 
as a surface of revolution of a semi-circle 
about a vertical axis as illustrated on Fig.1. 

The total curvature of the rim $c_m + c_p$ depends on 
the position along the
surface determined by the angle $\alpha$ (Fig.1). Indeed, one principal
curvature (meridional curvature) is constant,
\nz
$$c_m = {1\over\rho}\eqno(A1)$$
\nz
while the second one (parallel curvature) is given by
\nz
$$c_p={1\over\rho}\cdot{\cos\alpha\over{\cos\alpha+R/\rho}}\eqno(A2)$$
\nz
and changes on the rim.

The resulting dimensionless total curvature is
\nz
$$ J = {{2 \cos\alpha + R/\rho}\over{\cos\alpha + R/\rho}}\eqno(A3)$$
\nz
The element of the area of the rim is
\nz
$$dA_e = 2\pi\rho^2\Biggl({R\over\rho} + \cos\alpha\Biggr) d\alpha \eqno(A4)$$
\nz
and the total area of a circular rim is
\nz
$$A_e = 4\pi\rho^2(1 + {\pi\over 2}{R\over\rho})\eqno(A5)$$
\nz

Averaging the curvature and its square over the area of the rim gives
\nz
$$<J> = {{{\pi\over 2} + 2 {\rho\over R}}\over
 {{\pi\over 2} + {\rho\over R}}}\eqno(A6)$$
\nz
$$<J^2> =  {{2{\rho\over R}}\over {{\pi\over 2} + {\rho\over R}}}\cdot
\Biggl[2 + {1\over{{\rho\over R}\sqrt{1 -({\rho\over R})^2}}}
tg^{-1}\sqrt{{1-{\rho\over R}}\over{1+{\rho\over R}}}\Biggr]\eqno(A7)$$
\nz\nz\nz
\centerline{\bf Appendix B: Energy of phase of disc-like micelles.}
We consider the free energy of disc-like micelles. The 
shape of one micelle is illustrated on Fig.1. The area of monolayer
forming the rim of a micelle and that forming the flat central part are denoted
by $A_e$ and $A_f$, respectively. The composition of a micelle 
averaged over its entire area is $\phi_D$.

The free energy of a micelle is equal to the sum of the 
free energies of the rim and of the flat part.
Let us consider them separately.
\nz\nz
The free energy per unit area of the rim given by
(6) and accounting for  (3) and  (4) can be written in dimensionless form as
\nz
$$ f = (J - \zeta_L - \phi\cdot\Delta\zeta)^2 + \eta \phi (\phi - 1)\eqno(B1)$$
\nz
where $J$ and $\phi$ are, respectively, 
the local values of the curvature and composition.
We will add later the Gaussian curvature term as its integral over 
the micellar closed surface is equal to $8\pi {\bar\kappa /\kappa}$.

The composition of the rim averaged over its area will be denoted as
$<\phi> = \phi_e$. Minimizing the energy (A1) at fixed average composition
$\phi_e$ we find the distribution of the composition along the surface of the
rim
\nz
$$\phi=\phi_e + {{\Delta\zeta}\over{(\Delta\zeta)^2+\eta}}
\Biggl( J - <J> \Biggr)\eqno(B2)$$
\nz
Accordingly,
\nz
$$<\phi^2>=\phi_e^2 + {{(\Delta\zeta)^2}\over{[(\Delta\zeta)^2+\eta]^2}}
\Biggl(<J^2> - <J>^2 \Biggr)\eqno(B3)$$
\nz
and
\nz
$$<J\phi>=\phi_e<J> + {{(\Delta\zeta)}\over{(\Delta\zeta)^2+\eta}}
\Biggl(<J^2> - <J>^2 \Biggr)\eqno(B4)$$
\nz
Averaging the free energy (B1) over the area of the rim and accounting for the
equations (B2)-(B4) we obtain
\nz
$$ f_e = \zeta_L^2\quad +\quad <J^2> +\quad [(\Delta\zeta)^2+\eta]\phi_e^2
\quad-\quad
2 (\zeta_L+\Delta\zeta\cdot\phi_e)<J>\quad -$$
$$\quad -{{(\Delta\zeta)^2}\over{(\Delta\zeta)^2+\eta}}
\Biggl(<J^2> - <J>^2 \Biggr)\quad + \quad 
(2\zeta_L\Delta\zeta - \eta)\phi_e\eqno(B5)$$
\nz
The free energy per unit area for the flat central part of a micelle (6) in a
dimensionless form is
\nz
$$f_f = [(\Delta\zeta)^2+\eta]\phi_f^2 +
(2\zeta_L\Delta\zeta - \eta)\phi_f+\zeta_L^2\eqno(B6)$$
\nz
where $\phi_f$ is the composition of the flat part.

The total energy of a micelle is
\nz
$$F_D = f_e A_e + f_f A_f\eqno(B7)$$
\nz
while the compositions of the rim and the flat part satisfy the condition
\nz
$$\phi_e A_e + \phi_f A_f = \phi_D A_t\eqno(B8)$$
\nz
where $A_t = A_e + A_f$ is the total area of monolayer forming the micelle.
Minimizing the free energy (B7) and taking into account (B8) we obtain 
\nz
$$\phi_e = \phi_D + 
{\Delta\zeta\over {(\Delta\zeta)^2+\eta}}<J>{A_f\over {A_t}}\eqno(B9)$$
\nz
and
\nz
$$\phi_f = \phi_D -
{{\Delta\zeta}\over{(\Delta\zeta)^2+\eta}}<J>{A_e\over {A_t}}\eqno(B10)$$
\nz
The resulting expression for the free energy of a micelle is
\nz
$$ F_d = A_t\cdot f_D \eqno(B11)$$
\nz
where $f_D$ is the free energy of disc-like micelles per unit area 
given by (7).

The expressions for $<J>$, $<J^2>$, $A_e$ and $A_t$ for a specific model of 
micelle shape used here are derived in the Appendix A.
\nz\nz
\centerline{\bf Appendix C: Energy of phase of cylindrical micelles.}
Shape of a cylindrical micelle is assumed to consist of cylindrical body 
of length $l$
and area $A_l$ and two spherical caps of total area $A_h$. 
The radius of the cylinder and
the caps are assumed to be equal to $\rho$. Here, their area is equal to
$A_l = 2\pi\rho l$ and $A_h = 4\pi\rho^2$, respectively (Fig.2). 

Composition of a cylindrical
micelle averaged over its area will be denoted  as $\phi_c$, while
$\phi_h$ and $\phi_l$ are the compositions of the caps and cylindrical 
body of a micelle. In analogy with the calculation performed in Appendix B,
the dimensionless free energy of a micelle is
\nz 
$$ f_c = A_h \Biggl[ (2 - \zeta_L - \Delta\zeta\cdot\phi_h)^2 + \eta\phi_h^2 
- \eta\phi_h\Biggr] + 
A_l \Biggl[ (1 - \zeta_L - \Delta\zeta\cdot\phi_l)^2 + \eta\phi_l^2 
- \eta\phi_l\Biggr] + 8\pi {\bar\kappa\over\kappa}\rho^2 \eqno(C1)$$
\nz
The compositions $\phi_e$ and $\phi_l$ are related to the average composition
$\phi_c$ by
\nz
$$\phi_h A_h + \phi_l A_l = \phi_c A_t\eqno(C2)$$
\nz
Minimization of the energy (C1) while 
accounting for (C2) results in expressions
for the compositions of the caps and the cylindrical part
\nz
$$\phi_h = \phi_c + 
{\Delta\zeta\over {(\Delta\zeta)^2+\eta}}{A_l\over {A_t}}\eqno(C3)$$
\nz
$$\phi_l = \phi_c -
{{\Delta\zeta}\over{(\Delta\zeta)^2+\eta}}{A_h\over {A_t}}\eqno(C4)$$
\nz
Inserting (C3) and (C4) into (C1) we get the dimensionless free energy
of a cylinder micelle
\nz
$$ F_c = A_t f_c \eqno(C5)$$
\nz
where $f_c$ given by (8) is the free energy per unit area of surface of 
cylindrical micelles.

In analogy to the derivations above we obtain the expressions (9) and (10)
for the free energies of spherical micelles and flat bilayers.
\nz\nz
\centerline{\bf Appendix D: Derivation of compositions of phase transition}
\centerline{\bf by common tangent construction.}\nz\nz
The compositions of bilayers determining their transition to micelles 
satisfy the conditions
of equal $\partial f \over {\partial\phi}$ and equal 
$f - \phi {\partial f \over {\partial\phi}}$ in the both phases. Hence, 
the expression necessary to determine these compositions are
\nz
$${\partial f_D \over {\partial\phi_D}} =
2 [(\Delta\zeta)^2 + \eta]\phi_D +
(2\zeta_L\Delta\zeta - \eta - 2{A_e\over A_t}\Delta\zeta <J>)\eqno(D1)$$
\nz
$$ f_D - \phi_D {\partial f_D \over {\partial\phi_D}} =
- [(\Delta\zeta)^2 + \eta]\phi^2_D + \zeta_L^2 +$$
$$+ {A_e\over A_t}\Biggl[ {\eta\over{\eta+(\Delta\zeta)^2}}<J^2> +
{(\Delta\zeta)^2\over{\eta+(\Delta\zeta)^2}}<J>^2\Biggr] -$$
$$ - {A_e A_f\over A_t^2}{(\Delta\zeta)^2\over{\eta+(\Delta\zeta)^2}}<J>^2
- 2 {A_e\over A_t}\zeta_L<J> +
8\pi{\bar\kappa\over\kappa}{\rho^2\over A_t}\eqno(D2)$$
\nz
$${\partial f_c \over {\partial\phi_c}} =
2 [(\Delta\zeta)^2 + \eta]\phi_c +
2\zeta_L\Delta\zeta - \eta - 2\Delta\zeta 
- 2 \Delta \zeta {A_h\over A_t}\eqno(D3)$$
\nz
$$ f_c - \phi_c {\partial f_c \over {\partial\phi_c}} =
-[(\Delta\zeta)^2 + \eta]\phi^2_c  + (\zeta_L-1)^2 - 
{(\Delta\zeta)^2\over{[(\Delta\zeta)^2 + \eta]}}{{A_h A_l}\over A_t^2}+
(3-2\zeta_L){A_h\over A_t} + 8\pi{\bar\kappa\over\kappa}{\rho^2\over A_t}
\eqno(D4)$$
\nz
$${\partial f_s \over {\partial\phi_s}} =
2[(\Delta\zeta)^2 + \eta]\phi_s
+ (2\zeta_L\Delta\zeta - \eta - 4\Delta\zeta)\eqno(D4)$$
\nz
$$f_s - \phi_s {\partial f_s \over {\partial\phi_s}} =
- [(\Delta\zeta)^2 + \eta]\phi^2_s + 4
- 4\zeta_L + 2 {\bar\kappa\over\kappa} + \zeta_L^2 \eqno(D5)$$
\nz
$${\partial f_b \over {\partial\phi_b}} =
2 [(\Delta\zeta)^2 + \eta]\phi_b
+ (2\zeta_L\Delta\zeta - \eta) \eqno(D6)$$
\nz
$$f_b - \phi_b {\partial f_b \over {\partial\phi_b}} =
-[(\Delta\zeta)^2 + \eta]\phi_b^2 + \zeta_L^2\eqno(D6)$$
\vfill\eject
\centerline{\bf Figure captions.}
\nz\nz
Fig.1. Schematic  representation a disc-like micelle. 
$c_m = {1\over\rho}$ and $c_p$
are the meridional and parallel curvatures of the surface of
the  
rim, respectively, and $R$ is the radius of the flat central 
part of the micelle. Meridional angle $\alpha$ determines position along the
profile of the rim.
\nz\nz
Fig.2. Schematic representation of a cylindrical micelle. $l$ is the length 
of cylindrical body, $\rho$ is the radius of curvature of the 
semi-spherical caps.
\nz\nz
Fig.3. Free energies of the different states of the system.\nz
- liposomes (bilayers) (curve b, according to Eq.10),\nz 
- disc-like micelles (curve d, according to Eq.7),\nz
- cylindrical micelles (curve c, according to Eq.8),\nz 
- the phase of coexisting liposomes and 
disc-like micelles (common tangent b-d),
\nz
- the phase of coexisting liposomes and cylindrical micelles (common tangent b-c)
\nz
The chosen parameters on the curves are: 
$\eta = 0.75$, $\Delta\zeta = 1.$, $\zeta_L = 0$,
$\bar\kappa = 0$, the radius of a disc-like micelle $R = 100\cdot\rho$,
the length of a cylindrical micelle $l = 10000\cdot\rho $.
\nz
Points (1) and (2) on the $\phi$-axis indicate the critical compositions
$\phi^*(b\rightarrow d)$ and $\phi^*(b\rightarrow c)$, respectively.\nz\nz
Fig.4. Critical compositions of transition of liposomes into disc-like
micelles ( $\phi^*(b\rightarrow d)$, curve $d$ according to Eq.12) 
and into cylindrical micelles 
( $\phi^*(b\rightarrow c)$, curve $c$ 
according to Eq.11) as functions of surface area of one micelle
for different values of the parameter $\lambda$.
(For comparison, the dashed line $s$ (Eq.13) shows the constant critical
composition of transition of liposomes to spherical 
micelles ($\phi^*(b\rightarrow s)$).\nz
(A) $\lambda = -0.3$; (B) $\lambda = -0.6$; (C) $\lambda = -2.$;
\nz\nz
Fig.5 Phase diagram of the shapes of micelles resulting from solubilization of 
bilayers (liposomes) at different values of parameter $\lambda$.
\vfill\eject\end